# Atomic Scale Plasmonic Switch

*Alexandros Emboras[1]\*, Jens Niegemann[1], Ping Ma[1], Christian Haffner[1], Andreas Pedersen[2], Mathieu Luisier[2], Christian Hafner[1], Thomas Schimmel[3] and Juerg Leuthold[1]\*\**

[1] Institute of Electromagnetic Fields (IEF), ETH Zurich, 8092 Zurich, Switzerland.

[2] Computational Nanoelectronics Group, ETH Zurich, 8092 Zurich, Switzerland.

[3] Institute of Applied Physics and Institute of Nanotechnology (INT), Karlsruhe Institute of Technology (KIT), 76128 Karlsruhe, Germany

ABSTRACT: The atom sets an ultimate scaling limit to Moore's law in the electronics industry. And while electronics research already explores atomic scales devices, photonics research still deals with devices at the micrometer scale. Here we demonstrate that photonic scaling – similar to electronics - is only limited by the atom. More precisely, we introduce an electrically controlled plasmonic switch operating at the atomic scale. The switch allows for fast and reproducible switching by means of the relocation of an individual or at most – a few atoms in a plasmonic cavity. Depending on the location of the atom either of two distinct plasmonic cavity resonance states are supported. Experimental results show reversible digital optical switching with an extinction ratio of 9.2 dB and operation at room temperature up to MHz with femtojoule (fJ) power consumption for a single switch operation. This demonstration of an integrated quantum device allowing to control photons at the atomic level opens intriguing perspectives for a fully integrated and highly scalable chip platform - a platform where optics, electronics and memory may be controlled at the single-atom level.

Scaling devices towards atomic dimensions is a technological challenge, a challenge that however may pay off with devices operating at higher speed, lower power consumption and the highest possible device density[1]. Most of the research at atomic scale so far comes from electrical devices[2-11]. Of particular practical interest are the single atom transistor[5] and switches[6]. Similar to such technologies a non-volatile digital atomic switch based on the memristive technology platform have emerged[6,7]. The memrestive technology offers intriguing perspectives for the future development of the CMOS microelectronics industry because it can maintain its state after being activated and therefore can significantly decrease the switching energy[8]. The principle of operation is based on the electrically induced reversible motion of atomic scale species which shunts/opens two metallic electrodes through the insulator. Thus, the electrical resistance can be abruptly (digital) switched on/off and electrically read out.

Although electronic scaling reaches atomic scales dimensions, photonic scaling[12-14] still deals with devices with micrometer dimensions[15-21]. Recently, there has been quite some effort towards the realization of photonic systems that are sensitive to the presence of a single atom. The investigation of such systems is challenging because the atom-photon interaction is generally weak. One approach to face up this issue is to place a single atom in high finesse macroscopic cavities [22-25]. Based on such approaches a strong and controlled interaction between matter on the atomic scale and optical photons has been demonstrated [22-25]. All these demonstrations are important steps towards the development of new devices for quantum information processing. However, the manipulation of these atoms inside the cavities typically requires complicated and expensive experimental techniques and often the experiments only work by magneto-optic atom trapping at cryogenic temperatures which come along with challenging packaging issues.

Here we demonstrate an integrated optoelectronic quantum switch operating at the atomic level[26]. Depending on the quantum state of few if not a single atom, the switch is in either of two digital states. Step-like switching performance with 9.2 dB extinction ratio with as little as 12.5 fJ/bit and operation up to 1 MHz is reported. The switch shows reliable and robust performance at room temperature. The behavior agrees with *ab-initio* and FDTD (Lumerical Inc.) calculations that indicate how changing the conductive behavior by e.g. relocating a single atom may completely change the plasmonic resonances.

The operation principle and the configuration of the atomic scale switch is discussed with the help of Figure 1a. The switch consists of a planar silicon photonic waveguide with a plasmonic metal-insulator-metal (MIM) slot waveguide on top. The silicon waveguide (Si-WG) is used to launch the optical signal into the MIM slot waveguide where it is converted into a gap plasmon. The MIM slot waveguide is then adiabatically tapered down and ultimately guides the plasmon into a metallic tip with a separation distance that is as narrow as ~20 nm [27]. The plasmonic switching is then performed in the area formed between the metallic tip of the left metal and the right metal. The optical modulation is then controlled by applying a voltage between the two metals, Figure 1b, c. The process is described in more detail further below. But in essence, a single metallic nanofilament is formed at the tip that creates a plasmonic cavity. By moving a single or at most few atoms in the cavity the two metals are either short-circuited and thus conductive or not-short circuited and thus separated by a high ohmic resistance. The presence and absence of conductance alters the plasmonic cavity properties to the extent that two distinct plasmonic resonances are supported. It will be seen further down, that an enhanced conductance is obtained when the atom comes sufficiently close so that tunneling becomes possible.

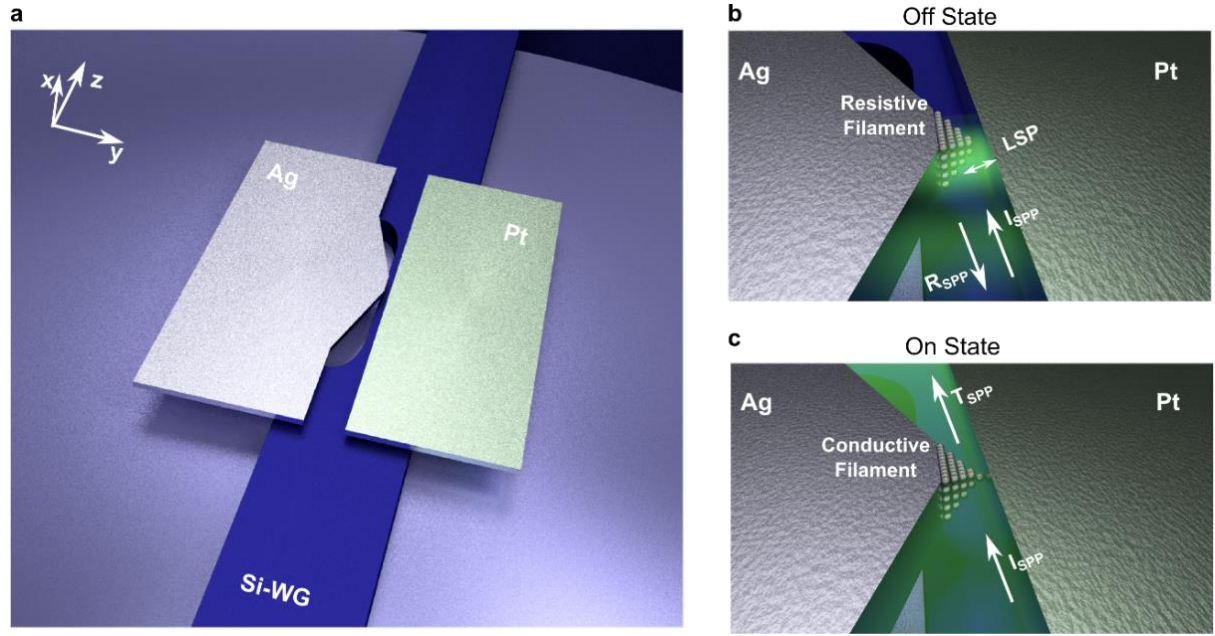

Figure 1. Schematic of Atomic scale plasmonic switch. a, 3D visualization of the atomic scale plasmonic switch coupled to a silicon waveguide. An optical signal is converted by means of the metallic pads to a gap plasmon and subsequently guided into the metallic tip where switching operation is performed. b, and c show a blow-up of the metallic tip comprising the nano-filament. By adding a single contact atom to the nano-filament the two metallic pads are short-circuited and the electrical conductance – and with this the resonance property of the cavity - dramatically changes. Since the two cavities have different resonance states one can observe a pronounced switching.

We fabricated an integrated atomic-scale optical switch on a 220 nm silicon-on-insulator (SOI) wafer. First, the silicon waveguides of 220 nm x 500 nm cross section were fabricated by using a modified local oxidation of silicon (LOCOS) approach which provides nearly planarized waveguides [28]. The Si-WG is then encapsulated with 50 nm of PECVD $Si_3N_4$. This is followed by the fabrication of the metallic silver (Ag) and platinum (Pt) pads by using two e-beam lithography steps and a subsequent lift-off process. Then, the slot between the two metals was filled with a-Si which acts as matrix for the silver ions. The exact geometry of the fabricated electro-optical device is given in the supplementary materials. Finally the atomic scale filament structure was fabricated by using a compatible memristive technology. More precisely, under the application of a positive voltage at the Ag electrode (the Pt electrode is grounded), silver ions become mobile and move into the aSi forming an atomic-scale filamentary channel by capturing free electrons coming from the Pt [29, 30], Figure 2a. It is worth noting that during this process the filament always forms at the spot where the separation distance between the two electrodes is minimum. Indeed, at this location is the maximum of the electric field where a filament preferentially develops. To support the assertion of a memrestive filament creation we perform a voltage-current characterization, black line in Figure 2b. The device reveals the typical resistive memory characteristics when the voltage is swept from -0.25V to 1.5 V [31]. The switching threshold voltage is around 1.25V. A compliance current has been set to 10 nA which means that the device switches with as little as 12.5 nW of power.

The plasmonic switching characteristics have subsequently been tested. At first a sinusoidal electrical input signal has been applied. Upon increasing the voltage, the optical signal increases abruptly (digitally) at around 1.25V. A plot with the sinusoidal electrical signal and the optical signal detected at the output is shown in Figure 2c. It is seen how the sinusoidal electrical input signal is converted into a signal with a digital optical switch performance. The optical transmission remains high until the voltage reaches a threshold voltage of around 0.25V from where the optical transmission is again abruptly attenuated. This is to the best of our knowledge the first time that such abrupt digital performance has been found for an optical switch. It is inherently related to the fact that the switch resides in either of two quantum states, exhibiting a digital rather than a continuous transition, in agreement with the mechanism of relocations on the level of individual atoms. To more clearly relate the optical switching to the conductance, the normalized optical output transmission (blue lines) of the devices was measured simultaneously to the current-voltage characteristic (black lines), see Figure 2b. The abrupt optical transition is observed when the switch becomes conductive and it switches off when the applied voltage is swept back to -0.25V.

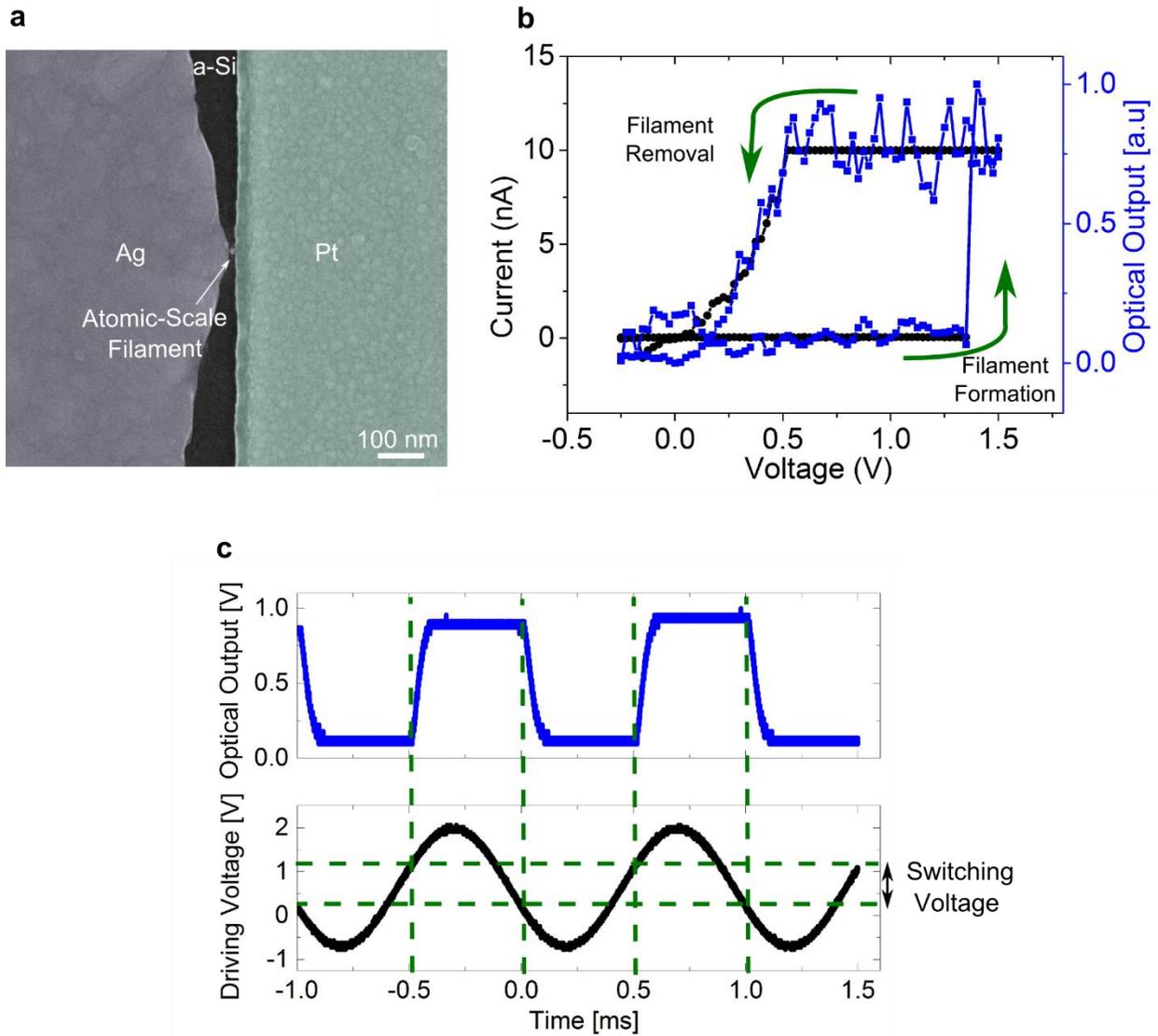

Figure 2. Two level quantum plasmonic switch. a, Top SEM image picture of the integrated atomic scale switch after formation of an atomic scale filament. b, Memristive current-voltage characteristic (black solid circles) and the corresponding output optical power (blue solid squares) performed at the same time. c, Switching performance as a function of the input voltage and time. A sinusoidal electrical input signal is fed to the device. A digital optical output power is found.

Operation at the atomic level can be asserted if conductance quantization and the conductance fingerprints of atomic-scale relocations are observed for a device in the "on-state". For this we first bring the device into the "on-state" and then measure the conductance during a current-voltage sweep – where we characterize the device for a voltage range that exceeds the normal operation conditions. In Figure 3a one can see how the device initially resides in a state with a low conductance of 0.13 $G_0$, i.e. a fraction of the fundamental quantum conductance unit ($G_0$=2e$^2$/h=7.74 × 10$^{-5}$ S) for up to a voltage of 1.8 V. When further increasing the voltage, the conductance changes to a value of a single quantum conductance unit. Sharp transitions with a clear and well defined conductance plateau are then observed, strongly resembling those observed in single atom devices[2-11].

To elucidate the operation mechanism and to show that single-atom operation is possible[32] we have performed *ab-initio* quantum transport calculations based on the density-functional theory (DFT) and the Non-equilibrium Green's function (NEGF) formalism[33]. Details about the simulation methodology and the creation of the Ag nano-filaments are provided in the Supporting Information. As shown in Figure 3b-c, the relocation of a single atom appears to be sufficient to change the filament conductance $G$ by several orders of magnitude and to go from the off-state ($G$=10$^{-5}$ $G_0$) to the on-state of the switch ($G$=0.13 $G_0$). By adding one more atom into the cavity, the conductance increases from 0.13 $G_0$ up to 0.96 $G_0$, see Figure 3d. Further conductance enhancements occur when more atoms are placed into the cavity, i.e. the conductance increases from 0.96 $G_0$ up to 1.5 $G_0$ when three instead of two atoms fill the gap between the filament and the right contact, see Figure 3e.

The simulations and experiments suggest that e.g. a 0.9 nm gap such as depicted in Figure 3b is large enough to guarantee a very low conductance that can be associated with the "off-state".

By adding a single atom into the narrow gap the filament head gets closer to the right contact and quantum mechanical tunneling sets in. The conductance abruptly increases from $10^{-5}$ to 0.13 $G_0$ and brings the switch in the "on-state", Figure 3c. By further increasing the applied voltage, e.g. a second atom is introduced in the cavity, as illustrated in Figure 3d. This way a conductive bridge is established between the filament and the right contact so that a single quantum channel becomes available ($G$=0.96 $G_0$). Finally, with three atoms in the gap ($G$=1.5 $G_0$, Figure 3e), a second quantum channel starts to form, but it is not completely open so that electrons must tunnel through a potential barrier to reach the right contact, as in the $G$=0.13 $G_0$ case. It should be notated, that other possibilities to explain the non-integer conductance quantum, e.g. back scattering of electron waves due to defects, impurities[4], or the Ag/Pt interface cannot yet be completely ruled out.

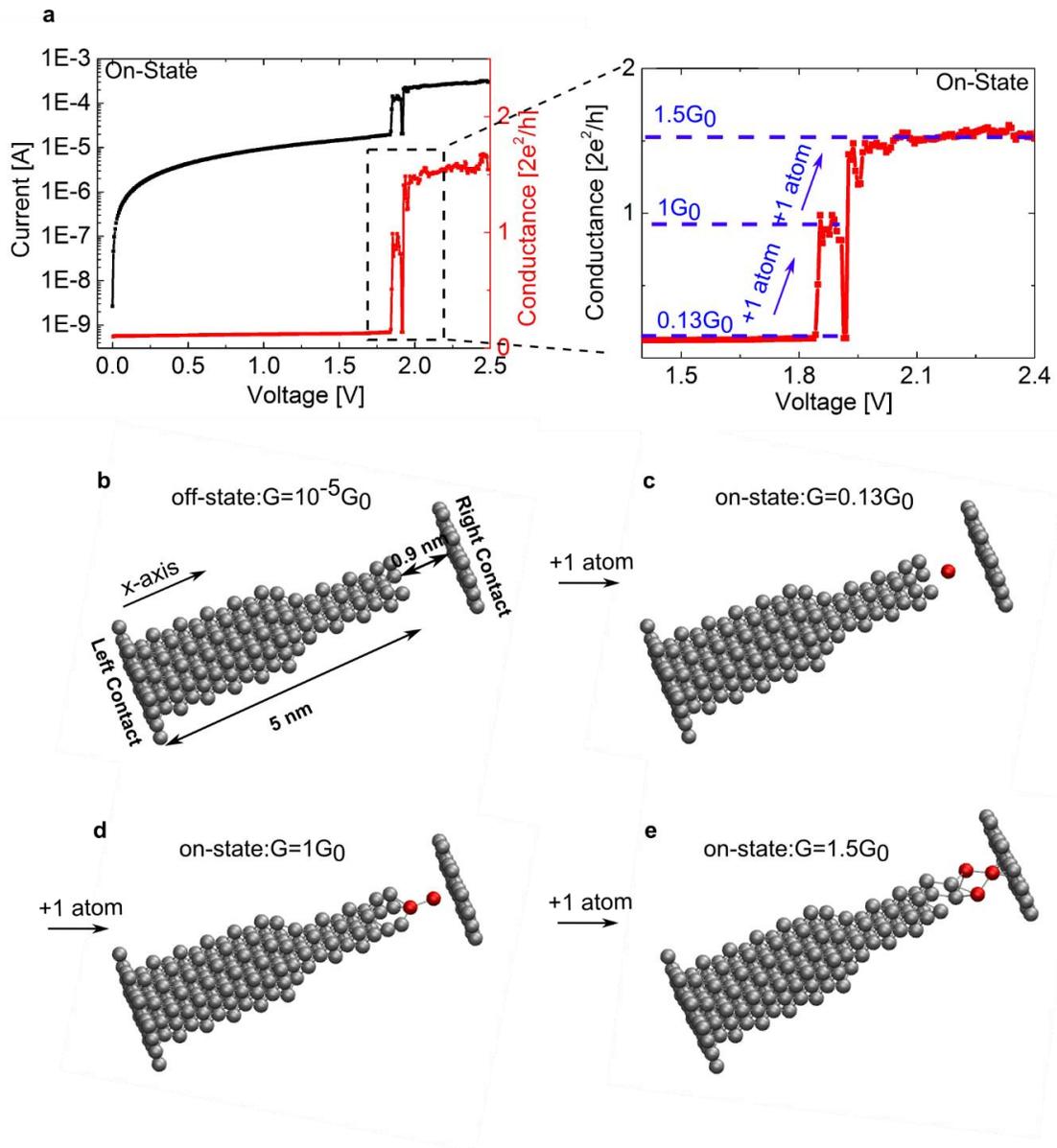

Figure 3. Indication of switching by atomic scale relocations. a, Current-voltage characteristic of the on-state (black curve) plotted in logarithmic scale. From this curve one can derive the conductance plot (red curve) that reveals quantized levels as a consequence of performing a current voltage sweep. Simulations indicate that these quantized states can be associated to different atomistic nano-filament arrangements as exemplarily depicted in (b-e). (b) The off-state corresponds to a filament that is sufficiently far offset from the right contact. (c) A single atom can open a conductive tunneling channel with a conductance in the order of e.g. 0.13 $G_0$. (d) Adding another atom can finally open single atom conductive channel. (e) With three atoms in the cavity another higher conductance level appears.

For a better understanding of the plasmonic switching principle of the device, we perform three-dimensional FDTD simulations (Lumerical Inc.). Figure 4a shows the calculated normalized transmission as a function of the wavelength for the "on" and "off" states. For the off state (red curve) the gap between the nanoscale filament and the platinum electrode is assumed to be 5 nm. In this case the transmission shows a broadband and pronounced minimum at the wavelength of 1.45 µm. This behavior is mainly attributed to the excitation of the plasmonic resonance localized between the nanoscale filament and the platinum electrode[34][35]. When switching the device from the off to the on state, the conductance abruptly changes due to, the relocation of a single or few atoms that now short-circuit the two pads by allowing for enhanced tunneling between the electrodes. As a consequence, the local field in the junction is modified[36,37] and this leads to a blue-shift and broadening of the resonance as can be seen in Figure 4a (black curve). Such an effect is expected when reducing the capacitive coupling between metallic electrodes. A top view (see Figure S1a and b for a plot of the exact structure) of the electric field intensity are depicted in Figure 4b for the case without and with filament. This behavior is analogous to recent findings where a blue shift of the resonance of the coupled plasmonic modes was observed when two neighboring plasmonic structures were bridged [38-42].

We experimentally measure the transmitted power as a function of the wavelength, see Figure 4b. The spectrum for the on and off states reveals a minimum in the transmission which is blue shifted for the on state, which is qualitatively in good agreement with the simulations. The shape and the positions of the resonances in the experiment do not completely agree with those of the calculations, which is to be expected because the precise geometric data of the experimental filament structure is not exactly known [30].

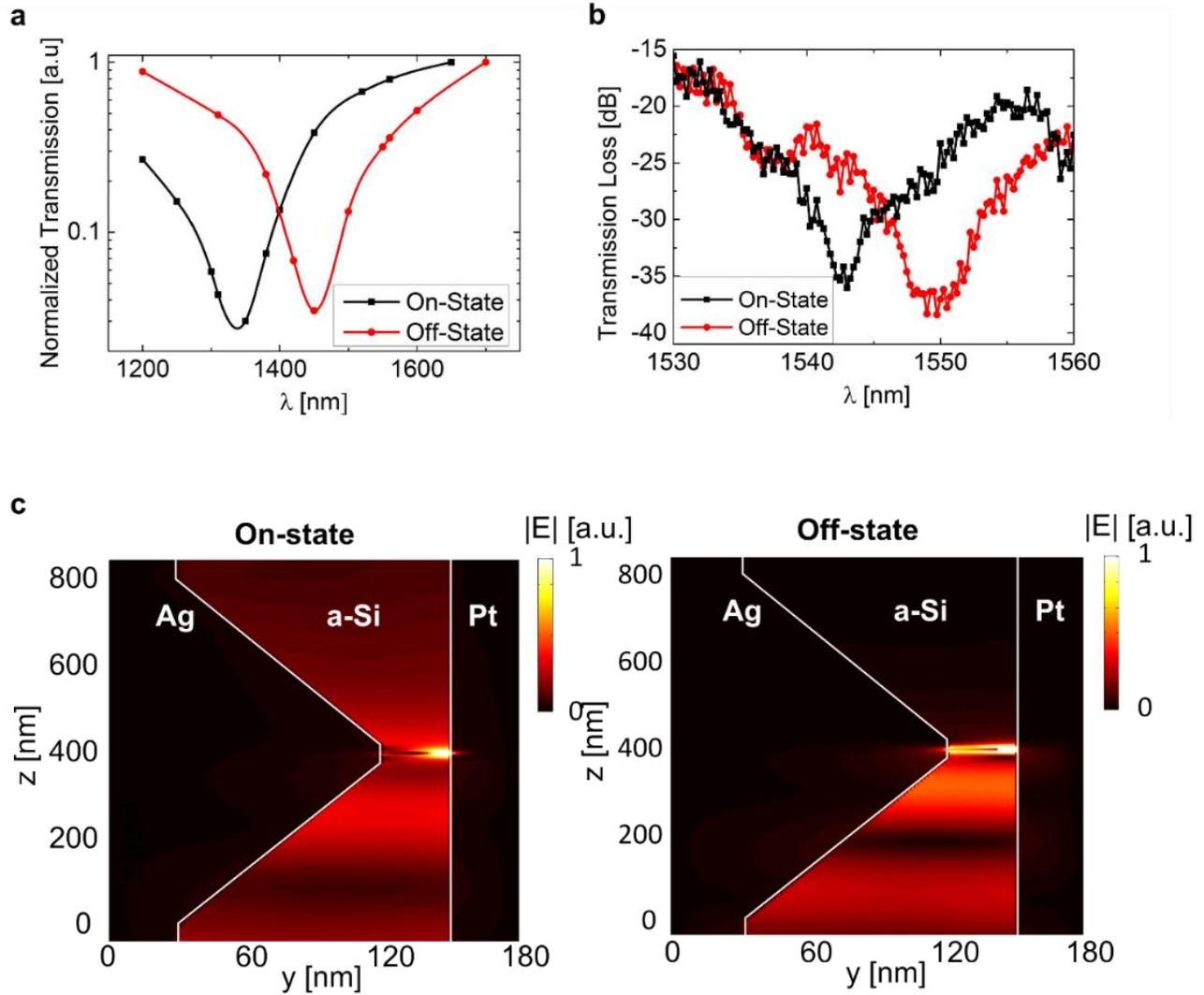

Figure 4. Switching mechanism and spectral response of switch in on- and off-state. a, Normalized transmission as a function of wavelength obtained by FDTD calculations. Two distinct resonances can be seen. b, Experimental transmission as a function of the wavelength. c, Top view of the E-field propagating along the waveguide (z-direction) direction (λ=1450 nm) for a device in on state (i.e. where an atom induces a short-circuit between the Ag and Pt electrodes) and a device in off state (where the atom is pulled back and the circuit is open).

To further investigate the switching speed, a small signal sinusoidal modulation was applied to the device and detected with a high speed photodiode and lock-in amplifier. The frequency response is flat from 50 kHz up to ~1MHz. It is somewhat enhanced by up to a factor 2 for slow modulation frequencies below 50 kHz (see Figure 5a). Higher switching rates may be anticipated in the future. For this the mobility of the silver atoms inside the switching medium has to be improved and the electrochemical potentials of the material system need to be adapted. Considering the fundamental limits of physics, frequencies up to 10s of GHz may be feasible for processes based on relocations of individual atoms[43]. On the basis of the numbers given from Figure 5 and with the help of Figure 2b one can also estimate the energy required for switching which is about 12.5 fJ/bit. Finally the durability of the electro-optical operation is conveniently assessed by the measurement of a large signal eye diagram Figure 5b. The eye diagram is a result of overlapping the output waveforms of a total of 20 574 symbols obtained from a pseudo random input bit sequence of length $2^{15}$-1 (500 µs of length for each symbol). The clear and wide eye opening are indicative for a reliable and error free switching way beyond many 1000s of switching cycles in all possible combinations of "0" and "1s". Also, we did not observe any degradation even when operating the device for many hours. The small fluctuations during the turn on and turn off transitions are attributed to resistance instabilities and the morphology of the voltage induced nanofilament[8].

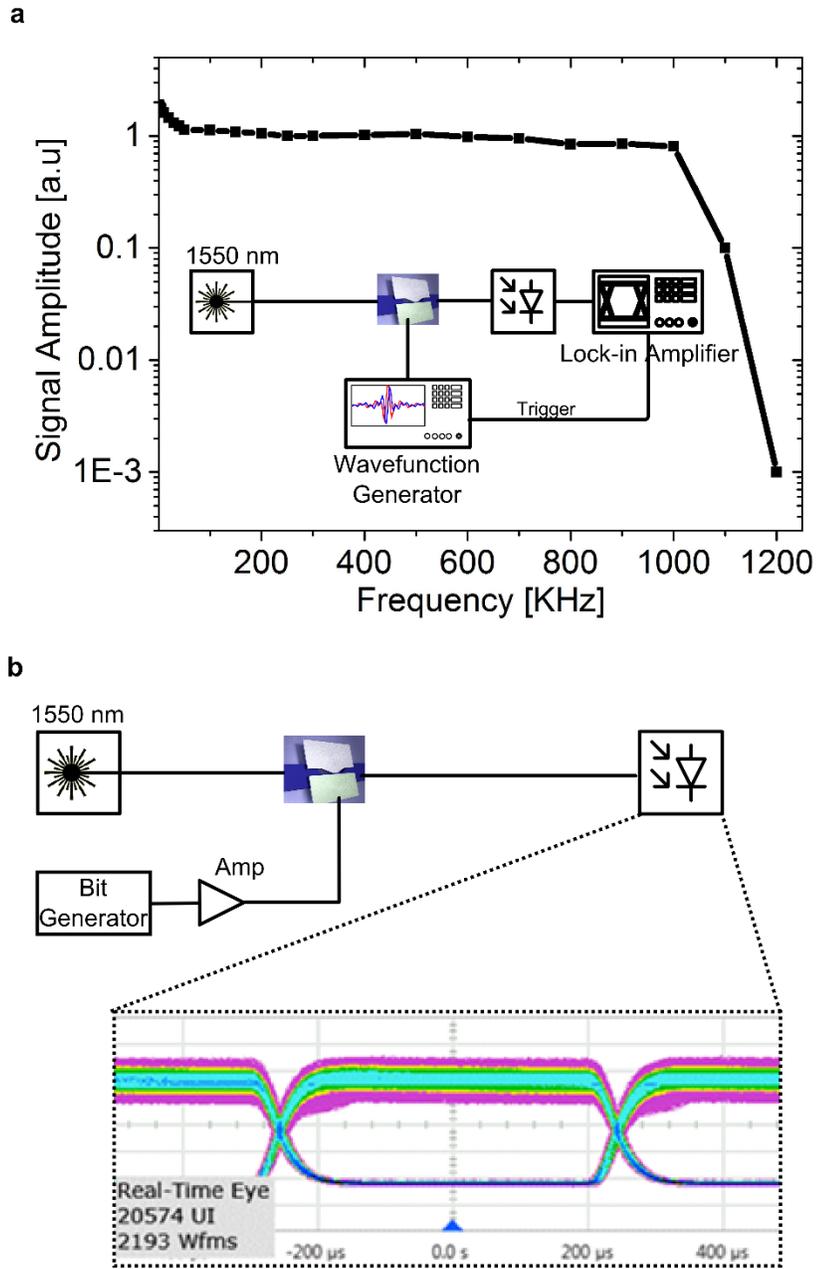

Figure 5. Frequency response. (a) Small signal frequency response: an electrical impulse was sent to the device under test (DUT) and the optical response was measured with a photodiode on a lock-in amplifier. Flat frequency response as a function of the applied sinusoidal RF-signal up to 1 MHz for an optical carrier at 1549 nm. (b) Large signal eye diagram measured by a real-time oscilloscope: A pseudo random bit sequence in the electrical domain is fed into the device and performs digital optical switching on a laser. The output signal was received by a photodiode and fed to an oscilloscope. An eye diagram is formed by overlaying several thousands of switching cycles.

In conclusion, we have introduced a novel photonic switch at atomic scale, performing digital operation with 9.2 dB extinction ratio and fJ power consumption per switching cycle. The switch might lay the foundation for a new integrated photonic atomic scale technology that works efficiently at ambient conditions. The device demonstrated here shows latching performance and thus might also give access to a new type of ultra-compact memory. It opens perspectives for combined electronic and plasmonic quantum devices at the single-atom level.

ASSOCIATED CONTENT

**Supporting Information**:

- Methods (Numerical Simulations, Device fabrication, Electro-Optical characterization)
- Supplementary Discussion

This material is available free of charge via the Internet at http://pubs.acs.org.

AUTHOR INFORMATION

**Corresponding Author**

*Email: aemboras@ethz.ch
**Email: leuthold@ethz.ch

**Notes**

The authors declare no competing financial interest

**Author Contributions**

A. Emboras conceived the concept, designed and fabricated the device, designed and performed the experiments and simulations, analyzed the data and wrote the paper. J. Niegemann conceived the concept and supported the project with simulations. P. Ma designed the experiments and wrote the paper. C. Haffner provided support in writing of the paper and preparing the figures. A. Pedersen and M. Luisier perform the *ab-initio* calculation and provided support in writing of the paper, C. Hafner, Th. Schimmel and J. Leuthold conceived the concept, designed the experiment, perform simulations and wrote the manuscript.

ACKNOWLEDGMENT

The Volkswagen Stiftung and ETH Zurich (Grant ETH-35 15-2) are acknowledged for partial funding. AP and ML would also like to thank the European Research Council under Grant Agreement No 335684-E-MOBILE for financial support. The work was carried out at the Binnig and Rohrer Nanotechnology Center, Switzerland. This work used computational resources from the Swiss National Supercomputing Centre under Project No. s579.